\documentclass[prd,twocolumn,twoside,preprintnumbers,superscriptaddress,nofootinbib,natbib,bm,amsrefs]{revtex4-2}
\usepackage{graphicx,subcaption,float}
\usepackage{hyperref}
\usepackage{url}
\usepackage{ragged2e}
\usepackage{xcolor}
\usepackage{amsmath, amstext, amsfonts,amssymb, amsthm}
\usepackage{comment}
\usepackage[utf8]{inputenc}
\usepackage{multirow}
\newcommand{\ov}{\overline}

\newcommand{\SM}{{\rm SM}}

\definecolor{BlueViolet}{rgb}{0.2, 0.00, 0.7}
\definecolor{Blue}{rgb}{0.15, 0.00, 0.9}
\definecolor{lightblue}{rgb}{0.15, 0.35, 0.95}
\definecolor{kitgreen}{rgb}{0,
0.58823 
, 0.50980 
}

\newcommand{\Eprint}[1]{\href{#1}}

\definecolor{lb}{rgb}{.74,.83,.9}
\definecolor{ly}{rgb}{1,.92,.8}
\definecolor{lr}{rgb}{.98,.85,.87}

\newcommand{\si}[1]{\textcolor{blue}{#1}}

\begin{document}
\preprint{ZU-TH 46/25}
\title{
Searching for Di-Higgs Signatures of Light Charged Scalars
}

\author{Guglielmo Coloretti}
\email{guglielmo.coloretti@physik.uzh.ch}
\affiliation{Physik-Institut, Universit\"{a}t Z\"{u}rich, Winterthurerstrasse 190, CH–8057 Z\"{u}rich, Switzerland}

\author{Andreas Crivellin}
\email{andreas.crivellin@cern.ch}
\affiliation{Physik-Institut, Universit\"{a}t Z\"{u}rich, Winterthurerstrasse 190, CH–8057 Z\"{u}rich, Switzerland}

\author{Syuhei Iguro}
\email{igurosyuhei@gmail.com}
\affiliation{Institute for Advanced Research, Nagoya University, Nagoya 464-8601, Japan}
\affiliation{Kobayashi-Maskawa Institute for the Origin of Particles and the Universe, Nagoya University, Nagoya 464-8602, Japan}

\begin{abstract}
The excess in $t\to b\overline{b}c$ observed by ATLAS points towards a charged Higgs boson with a mass around 130\,GeV, consistent with the expectations from the $B$ anomalies, i.e.~$R_{D^{(*)}}$ and $b\to s\ell^+\ell^-$ data. As a non-minimal flavour structure is required for an explanation of these observables, this points towards a two-Higgs-doublet model with generic Yukawa couplings. Such a scenario predicts a sizable cross section for the pair production of the charged Higgs at the Large Hadron Collider, which can be tested by recasting SM di-Higgs searches. While the predicted event rate is even higher than the one of SM Higgs pair production, the smaller efficiency (w.r.t.~SM Higgs pair production) reduces the signal yield. Nonetheless, dedicated searches can probe most of the interesting parameter space and lead to a discovery with Run-3 or High-Luminosity LHC data.\\
---------------------------------------------------------------------------------------------------------------------------------
\end{abstract}
\maketitle

\section{Introduction}

The Brout-Englert-Higgs boson~\cite{Higgs:1964ia,Englert:1964et,Higgs:1964pj,Guralnik:1964eu} had been the last missing piece of the Standard Model (SM)~\cite{Glashow:1961tr,Weinberg:1967tq,Salam:1968rm} until its discovery at the Large Hadron Collider (LHC)~\cite{ATLAS:2012yve,CMS:2012qbp} at CERN in 2012. While the couplings of this scalar to fermions and gauge bosons are quite well known and are in agreement with the SM predictions~\cite{CMS:2022dwd,ATLAS:2022vkf}, the structure of the Higgs potential still needs to be confirmed experimentally. In particular, the tri-linear Higgs coupling can be determined via Higgs pair production. While the current LHC analyses still have statistical uncertainties of order one~\cite{CMS:2024ymd,ATLAS:2024ish}, more precise determinations are among the main targets of the high luminosity (HL)-LHC program~\cite{Apollinari:2015wtw,Dainese:2019rgk} and a major motivation for future $e^+e^-$ colliders~\cite{FCC2018,CLIC2018,ILC2022,CEPC2018}.

Importantly, the observation that the measured properties of the SM(-like) Higgs boson are consistent with expectations does not exclude the existence of additional scalars, even at the electroweak (EW) scale, if the mixing is small. In fact, statistically significant indications for new EW scale Higgs bosons have emerged~\cite{LEPWorkingGroupforHiggsbosonsearches:2003ing,CMS:2018cyk,CMS:2022rbd,CMS:2022tgk,ATLAS:2023jzc,Crivellin:2021ubm,Bhattacharya:2023lmu,Banik:2023vxa,Coloretti:2023yyq,Crivellin:2024uhc,Ashanujjaman:2024lnr}, including an excess ($3\sigma$) in $t\to b H^+$ with $H^+\to \bar b c$ and $m_{H^\pm}\approx130\,$GeV~\cite{ATLAS:2023bzb}. 

In this letter, we point out an interesting and novel relation between searches for the pair-production of SM Higgses, and effects of light charged Higgses in the $\bar b c$ channel, as suggested by the ATLAS excess: This (hypothetical) Higgs boson can be pair-produced at the LHC. Due to the limited energy resolution of ATLAS and CMS detectors and the finite mis-tag probability of a charm quark as a bottom one, this results in signatures overlapping with the search for the non-resonant productions of SM Higgs bosons. Therefore, these analyses can be recasted and adapted to scrutinize the 130\,GeV excess. 

While most of our analysis will not depend on the details of the model, we will consider for concreteness the two-Higgs-doublet model with generic Yukawa couplings (G2HDM). This model cannot only account for the 130\,GeV excess~\cite{Crivellin:2023sig},\footnote{See Refs.~\cite{Akeroyd:2022ouy,Bernal:2023aai,Arhrib:2024sfg} for alternative models explaining the ATLAS excess in $t\to b \bar b c$.} but is also particularly interesting in light of the so-called $B$-anomalies~\cite{Capdevila:2023yhq} where a light charged Higgs boson can explain the deviations from the SM predictions in $R_{D^{(*)}}$~\cite{HFLAV2024} and the $b\to s\ell^+\ell^-$ anomalies~\cite{Alguero:2023jeh,Hurth:2023jwr}. In fact, the global fit of this model to data~\cite{Athron:2024rir} predicts that $H^+ \to \bar bc$ is the dominant decay mode of the charged Higgs as suggested by the ATLAS excess.

\section{Model and observables}

A light charged Higgs boson ($H^\pm$)\footnote{A light scalar is also motivated by scenarios of spontaneous $CP$-violation~\cite{Nierste:2019fbx}.} can be accommodated in the generic two-Higgs doublet model (G2HDM)~\cite{Hou:1991un,Chang:1993kw,Liu:1987ng,Cheng:1987rs,Savage:1991qh,Antaramian:1992ya,Hall:1993ca,Luke:1993cy,Atwood:1995ud,Atwood:1996vj,Botella:2015hoa, Herrero-Garcia:2016uab} without a $Z_2$ symmetry. It has been shown that this model can account for the $B$ anomalies~\cite{Crivellin:2012ye,Crivellin:2013wna,Cline:2015lqp,Crivellin:2015hha,Lee:2017kbi,Iguro:2017ysu,Martinez:2018ynq,Fraser:2018aqj,Athron:2021auq,Iguro:2022uzz,Blanke:2022pjy,Ezzat:2022gpk,Fedele:2022iib,Das:2023gfz,Iguro:2018qzf,Iguro:2018fni,Crivellin:2019dun,Kumar:2022rcf,Iguro:2023jju,Crivellin:2023sig,Athron:2024rir}, both the charged ($R_{D^{(*)}}$) and the neutral current ones ($b\to s\ell^+\ell^-$), and the $\approx$\,130\,GeV excess in $t\to b\bar bc$~\cite{ATLAS:2023bzb}.

\begin{figure}[t]
    \centering
    \begin{subcaptionbox}{\label{fig:1}}[0.42\linewidth]
        {\includegraphics[width=\linewidth]{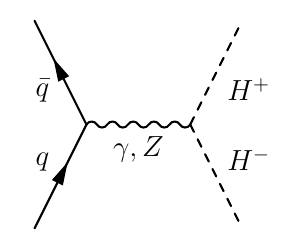}}
    \end{subcaptionbox}
    ~~
    \begin{subcaptionbox}{\label{fig:2}}[0.42\linewidth]
        {\includegraphics[width=\linewidth]{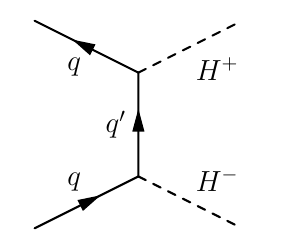}}
    \end{subcaptionbox}
    \vspace{-0.5cm}
    \begin{subcaptionbox}{\label{fig:3}}[0.42\linewidth]
        {\includegraphics[width=\linewidth]{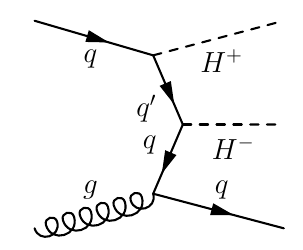}}
    \end{subcaptionbox}
    ~~
    \begin{subcaptionbox}{\label{fig:4}}[0.42\linewidth]
        {\includegraphics[width=\linewidth]{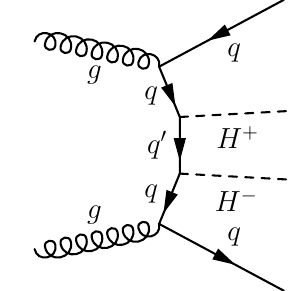}}
    \end{subcaptionbox}
    \vspace{-0.5cm}
    \begin{subcaptionbox}{\label{fig:5}}[0.42\linewidth]
         {\includegraphics[width=\linewidth]{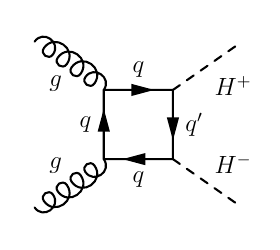}}
    \end{subcaptionbox}
    ~~
    \begin{subcaptionbox}{\label{fig:6}}[0.42\linewidth]
        {\includegraphics[width=\linewidth]{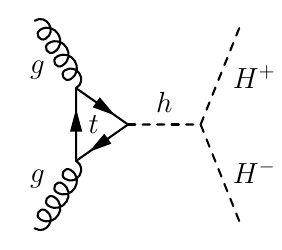}}
    \end{subcaptionbox}
    \vspace{4mm}
\caption{\justifying Feynman diagrams illustrating the leading contributions to charged-Higgs pair production at the LHC. }
    \label{fig:Fdiag}
\end{figure}

To perform our analysis, we will parametrize the Higgs couplings in the G2HDM in the so-called Higgs basis~\cite{Davidson:2005cw}, and consider a $CP$-conserving scalar potential for simplicity. In the Higgs basis, exploiting a residual global $U(2)$ symmetry of the scalar potential, only one Higgs doublet acquires a nonzero vacuum expectation value
\begin{eqnarray}
  H_1 =\left(
  \begin{array}{c}
    G^+\\
    \frac{v+\phi_1+iG^0}{\sqrt{2}}
  \end{array}
  \right),~~~
  H_2=\left(
  \begin{array}{c}
    H^+\\
    \frac{\phi_2+iA}{\sqrt{2}}
  \end{array}
  \right),
\label{HiggsBasis}
\end{eqnarray}
such that $G^+$ and $G^0$ are the would-be Goldstone bosons, and $H^+$ and $A$ are the charged Higgs and the $CP$-odd Higgs boson, respectively, and $v\approx 246\,$GeV. While $\phi_1$ and $\phi_2$ in general mix to result in the SM-like Higgs and an additional $CP$-even neutral scalar $H$, the neutral Higgses are of little consequence for our analysis\footnote{Note that as long as the production cross section of $H$ is sufficiently smaller than the one of the SM Higgs, its effect in the pair production of the charged scalars can be neglected if it cannot decay to them on-shell.} as long as the related constraints are satisfied~\cite{Crivellin:2023sig}.

While there are in general 9 additional Yukawa couplings (which are independent of the fermion masses) for each up-type quarks, down-type quarks and charged leptons, we will focus on the minimal set of Yukawa couplings necessary to explain the $B$-anomalies which are at the same time relevant for the production of charged Higgs pairs.\footnote{Note that $\rho_f^{ij}$, with $f=u,d,\ell$, is independent of the fermion masses. The off-diagonal elements of $\rho_d^{ij}$ are stringently constrained by meson mixing and decays. For a more detailed phenomenological analysis, see Refs.\,\cite{Crivellin:2013wna,Iguro:2017ysu,Iguro:2019zlc}.} Working directly with the Yukawa couplings of the charged Higgs $H^\pm$, we thus consider
\begin{align}
\mathcal{L}_Y = \; & \rho_u^{tc*}V_{td_i} \ov c P_L d_iH^+ +\rho_u^{tt*}V_{td_i} \ov t P_L d_iH^+ \\\nonumber 
&+\rho_u^{cc*}V_{cd_i} \ov c P_L d_iH^+ -\rho_\ell^{\tau\tau}\ov\nu_{\tau} P_R\tau H^+   +h.c.,
\end{align}
where $i$ is a flavor index and $V$ the Cabbibo-Kobayashi-Maskawa (CKM) matrix~\cite{Cabibbo:1963yz,Kobayashi:1973fv}.\footnote{In this setup, $m_A,\,m_H\ge m_t+m_c$ is required to avoid the LHC constraint from searches for multi-$\tau$ final state~\cite{Iguro:2023tbk}.}

\subsection{Observables}

Let us now look at the relevant observables in more detail. First of all, to explain the excess in the exotic top decay, we can fix $m_{H^\pm}=130\,$GeV and obtain
\begin{equation}
  \operatorname{Br}\left(t \rightarrow b H^{+}\right)=\frac{m_t\left|\rho_u^{t t}\right|^2}{16 \pi \Gamma_t}\!\left(1-\frac{m_{H^{\pm}}^2}{m_t^2}\right)^2 \!\!\approx 0.16\frac{\left|\rho_u^{t t}\right|^2}{0.06^2} \%  \,.
\end{equation}
This expression has been normalised such that the best-fit value 
\begin{equation}
    \operatorname{Br}\left(t\to b H^{+} \right)\times\operatorname{Br}\left(H^{+} \rightarrow \bar{b} c\right)=(0.16 \pm 0.06) \%\,,
\end{equation}
of the ATLAS excess is obtained for $\left|\rho_u^{t t}\right|=0.06$ if $H^\pm$ decays dominantly to $cb$. Note that the analogous CMS result is based only on Run-1 data~\cite{CMS:2018dzl} and is therefore not competitive in sensitivity. There is no excess in analogous searches with $\bar s c$~\cite{CMS:2020osd} and $\bar \tau \nu$~\cite{ATLAS:2024hya} final states, leading to the limits Br$(t\to bH^+ )\times$Br$(H^+\to \bar{s}c)\lesssim3\times 10^{-3}$ and Br$(t\to b H^+ )\times$Br$(H^+\to \ov{\tau}\nu)\lesssim4\times 10^{-4}$, respectively, for $m_{H^\pm}=130$\,GeV.

Concerning the $B$ anomalies, we have the indications for lepton flavour universality violation in the ratios $R_{D^{(*)}}$. Here, the Wilson coefficient scalar operator at the B meson mass scale, $O_{S_L} = (\overline{c}  P_Lb)(\overline{\tau} P_L \nu_{\tau})$ is generated:
\begin{equation}
C_{S_L}^{\tau}(\mu_b)
\approx 0.67\,\frac{\rho_u^{t c *} \rho_{\ell}^{\tau\tau *}}{0.1^2}\left(\frac{130 \,\mathrm{GeV}}{m_{H^{\pm}}}\right)^2 
\end{equation}
with the preferred value $C_{S_L}^{\tau}\approx -0.29 + 0.7 i$.\footnote{Note that polarization observables require that the NP effect must be (mainly) related to tau leptons~\cite{Fedele:2023ewe}.}

Secondly, there are the indications for new physics in neutral current $b\to s\ell^+\ell^-$ transitions. This includes $P_5^\prime$~\cite{Descotes-Genon:2013vna} and deficits in the total Branching ratios of other $b\to s \mu\ov\mu$ decays, Br$(B\to K\mu\ov\mu)$ \cite{LHCb:2014cxe,LHCb:2016ykl,Parrott:2022zte} and Br$(B_s\to \phi \mu\ov\mu)$, together with angular observables in the latter)~\cite{LHCb:2021zwz,Parrott:2022rgu,Gubernari:2022hxn}, as well as semi-inclusive observables~\cite{Isidori:2023unk}. The global fit currently prefers $C_9^U\approx -1$ where the operator of  $O_{9} = (\overline{s} \gamma_\mu P_L b)(\overline{\ell} \gamma^\mu \nu_{\ell})$ is defined.

Here, our NP effect proceeds via a charm loop induced off-shell photon penguin\footnote{In fact, given the updated tests of $\mu$-$e$ lepton flavor universality encoded in $R(K^{(\ast)})$~\cite{LHCb:2022qnv} and the $Q$ observables~\cite{LHCb:2025pxz}, which agree with the SM predictions, the current situation indicates a lepton-flavor-universal effect in $C_9^{\text{U}}=C_9^e=C_9^\mu\simeq-1$~\cite{Alguero:2023jeh,Hurth:2023jwr} as predicted in our setup.} resulting in 
\begin{equation}
    \Delta C_9^U\left(\mu_b\right) \approx-0.52\frac{\left|\rho_u^{t c}\right|^2-\left|\rho_u^{c c}\right|^2}{0.5^2}+0.50\frac{\rho_u^{t c *} \rho_u^{c c}}{0.1^2}\,.
\end{equation}
From these two anomalies, it is clear that $\rho_u^{tc}$ and $\rho_u^{cc}$ are, in fact, expected to be the numerically largest couplings (also in light of the constraints in the lepton couplings from LHC searches).

$b\to s\gamma$ and $B_s-\bar B_s$ mixing give relevant constraints on $\rho_u^{tc}$ and $\rho_u^{tt}$. Adopting the global fit of Ref.~\cite{Wen:2023pfq} we find $-0.035\lesssim{\rm{Re}} [\Delta C_7(\mu_b)]\lesssim 0.037$ at the $2\sigma$ level. We obtain the semi-analytic formula for one-loop charged Higgs contribution \cite{Crivellin:2019dun}
\begin{align}
   \Delta C_7(\mu_b)\approx&-3\left(\rho_u^{cc}-0.04\rho_u^{tc}\right)\left(0.04\rho_u^{cc}+\rho_u^{tc}\right)^*\notag \\ 
   &-0.03|\rho_u^{tt}|^2,
   \label{eq:C7}
\end{align}
meaning that the G2HDM interferes constructively with the SM.
Using the input of the online update of Ref.~\cite{UTfit:2007eik} we obtain the allowed range of $-0.09\le R_{B_s}\equiv\Delta M_{B_s}^{\rm{G2HDM}}/\Delta M_{B_s}^{\rm{SM}}\le0.07$ \cite{Crivellin:2023saq}.
This has to be compared to
\begin{align}
R_{B_s}\approx 0.05\left|\frac{\rho_u^{tc}}{0.5}\right|^2\!\!+0.002\left|\frac{\rho_u^{tt}}{0.05}\right|^2\!\!-0.01\biggl|\frac{\rho_u^{tc*} \rho_u^{tt}}{0.025}\biggl|\,,
\label{eq:Bsmix}
\end{align}
for $m_{H^\pm}=130$\,GeV.\footnote{Here we omitted $\mathcal{O}\left((\rho^{ij}_u)^4\right)$ terms and the analogous formula from Kaon and $B_d-\bar B_d$ mixing. See Ref.~\cite{Crivellin:2023sig} for details.}
Also constraints from $K^0-\bar K^0$ mixing provide the relevant constraint \cite{UTfit:2007eik}.

\section{Di-Higgs signal strength}

The ATLAS excess contains the final state $b\bar b c$, and the charged Higgs mass is reconstructed from the invariant mass of the $\bar b c$ (or $\bar c b$) pair. Since there is no excess in analogous searches with $ s c$~\cite{CMS:2020osd} and $\tau \nu$~\cite{ATLAS:2024hya} final states, this implies that $ b c$ is the dominant decay rate while the $ \tau\nu$ mode is numerically smaller. Interestingly, as we have seen in the last section, this coupling pattern is consistent with the one preferred by a simultaneous explanation of the $B$ anomalies, i.e.~by $b\to c\tau\ov\nu$ and $b\to s\ell^+\ell^-$ data.

We now analyse the pair production of $H^\pm$ with the subsequent decay $H^\pm\to bc $.\footnote{Note that even if the decay mode $sc$ is sizable, as allowed by current limits, it is not important for our analysis since the probability of mistagging a strange quark as a bottom quark is small. Thus, only the branching ratio of the charged Higgs to $\bar cb$ is relevant.} There are several numerically relevant ways to produce two (hypothetical) light charged Higgs bosons at the LHC. In Fig.~\ref{fig:Fdiag}, Drell-Yan production through $Z^*$ and $\gamma^*$ (diagram ($a$)), as well as the tree-level production via the Yukawa couplings $\rho_u^{tc}$ and $\rho_u^{cc}$ (diagram ($b$)--($d$)), as well as at loop-level production via gluon fusion (diagram ($e$) and ($f$)) are shown. \footnote{Since $\rho_u^{tt}$ is numerically small in our scenario, and its effect in charged Higgs production is suppressed by the involved heavy top quark, we will only study $\rho_u^{tc}$ and $\rho_u^{cc}$ for the production cross section. Note that the cross section can be initiated via quark and gluon PDFs, resulting in zero, one or two additional forward jets. However, the efficiency in the vector-boson fusion channel (i.e., two forward jets) turns out not to be higher than the SM, such that this channel does not lead to competitive constraints compared to the gluon fusion signal region (as in the SM).} 

We will now use the analysis of non-resonant SM Higgs-pair production in the 4$b$ channel of ATLAS~\cite{ATLAS:2023qzf} to put bounds on the pair production of two charged Higgses. This is possible due to the mis-tag rate of a $c$-jet as a $b$-jet of around $20\%$ and because the typical jet energy resolution is larger than the mass difference between the charged Higgs and the SM Higgs of $\approx 5$\,GeV. Both SM and NP events are generated using \texttt{MadGraph5\_aMC\_v3.5.3}~\cite{Alwall:2014hca,Frederix:2018nkq} with the \texttt{NNPDF23\_nlo\_as\_0118\_qed} parton distribution function (PDF)\footnote{Note that the $H^\pm$ pair production cross section increases by $15$-$20\%$ at $\sqrt{s}=14$\,TeV. Similarly, the SM production cross section will increase by $18\%$~\cite{Borowka:2016ehy,Grazzini:2018bsd}.} using the pre-implemented \texttt{2HDM\_NLO}~\cite{G2HDMNLO} model file built with \texttt{Feynrules 2.0}~\cite{Alloul:2013bka}. The generated parton-level events are then passed through \texttt{Pythia 8.3}~\cite{Sjostrand:2014zea} for showering and \texttt{Delphes 3.5.0}~\cite{deFavereau:2013fsa} for the detector simulation. Object reconstruction and selection follow the ATLAS analysis~\cite{ATLAS:2023qzf}, with jets clustered using the anti-$k_T$ algorithm~\cite{Cacciari:2008gp} implemented in \texttt{FastJet 3.3.4}~\cite{Cacciari:2011ma}. Events are finally selected and categorized into the signal regions as defined in Ref.~\cite{ATLAS:2023qzf}.

The sum of the vector-boson fusion and gluon-fusion categories signal yields in our model, relative to the number of SM di-Higgs events in the ATLAS analysis~\cite{ATLAS:2023qzf}, is
\begin{widetext}
\begin{align}
    \mu_{4b} \approx 1+& \frac{\epsilon_{\rm SR}}{\sigma_{\rm SM}}\left(\frac{\epsilon_{c\to b}}{\epsilon_b}\right)^2 
     \biggl[
        \epsilon_{\rm Loop} \left(0.9 |\rho_u^{cc}|^4+0.2 |\rho_u^{cc}|^2 Z_3+0.01 Z_3^2+0.01 Z_3 |\rho_u^{tc}|^2+1 |\rho_u^{tc}|^4\right)\\
    &    + \epsilon_{\rm Tree} \left(6 |\rho_u^{cc}|^4+4 |\rho_u^{cc}\rho_u^{tc*}|^2-0.3 |\rho_u^{cc}|^2+0.02 \rm{Re}[\rho_u^{cc} \rho_u^{tc*}]+3 |\rho_u^{tc}|^4-0.06 |\rho_u^{tc}|^2+0.2\right)
    \biggl] \, {\rm pb}\times {\rm Br}(H^+\to \bar bc)^2\,,\notag
\end{align}
\end{widetext}
where we introduced the relative efficiencies (w.r.t.~the SM signal)
$\epsilon_{\rm Loop}\approx0.2$,  $\epsilon_{\rm Tree}=0.4$, $\epsilon_{\rm SR}\approx 1/2$, and the SM cross section $\sigma_\SM=\sigma_{hh}\times {\rm{Br}}(h\to b\ov{b})^2 \approx 9$fb. ''Loop`` and ''Tree`` refer to the level in perturbation theory at which the diagrams in Fig.~\ref{fig:Fdiag} appear. The probability of mistagging a charm quark as a bottom quark is $\epsilon_{c\to b}=0.2$ and the $b$-tag efficiency is $\epsilon_b=0.8$.\footnote{More precisely, the $b$-jet tagging efficiency is defined via $\epsilon_b \times {30 \, \tanh(0.003 \, p_T)}/({1+0.086 \, p_T}))$ while $
\epsilon_{c\to b}\times {\tanh(0.02 \, p_T)}/({1+0.0034 \, p_T})$
is the $c\to b$-mistagging rate, where $p_T$ in GeV is the transverse momentum of the jet as implemented in {\tt DELPHES}~\cite{deFavereau:2013fsa}.} $\epsilon_{\rm SR}$ accounts for the relative efficiency based on the number of events within the signal region in the plane of the two invariant di-jet masses, as defined by ATLAS.\footnote{For this, we corrected for the shift in the values of the di-jet masses due to the limitations of our fast simulation by shifting the ATLAS values to the mean values of our SM simulation.} $Z_3$ is defined via the Lagrangian term $L\supset-Z_3(H_1^\dagger H_1)(H_2^\dagger H_2)$. Note that the related contribution is small and thus will be neglected in the numerical analysis.

The observed (expected) $95\%$ CL upper limit on $\mu_{4b}$ from LHC Run-2 data is 5.4 (8.1) and 7.5 (4.3) from ATLAS~\cite{ATLAS:2023qzf} and CMS~\cite{CMS:2024ymd}, respectively. The tighter expected limit from CMS is due to the improved boosted bottom-jet reconstruction technique and the use of a boosted decision tree, which, however, makes recasting the analysis very difficult (therefore, we focused on the ATLAS analysis above). According to the ATLAS baseline projection, the HL-LHC with 1ab$^{-1}$ is expected to reach $\mu_{4b}\le 2.8$ at $2\sigma$ assuming a SM-like signal.\footnote{The total luminosity of 1ab$^{-1}$ is approximately the sum of the ones expected by ATLAS and CMS, and thus corresponds to the ``full'' Run-3 data set.} Furthermore, the ATLAS-CMS combined HL-LHC analysis will reach a significance of 2.8$\sigma$ for the SM signal. This roughly corresponds to a sensitivity of $\mu_{4b}\le 1.73$ at $95\%$ CL.

\begin{figure*}[t]
 \includegraphics[width=0.415\textwidth]{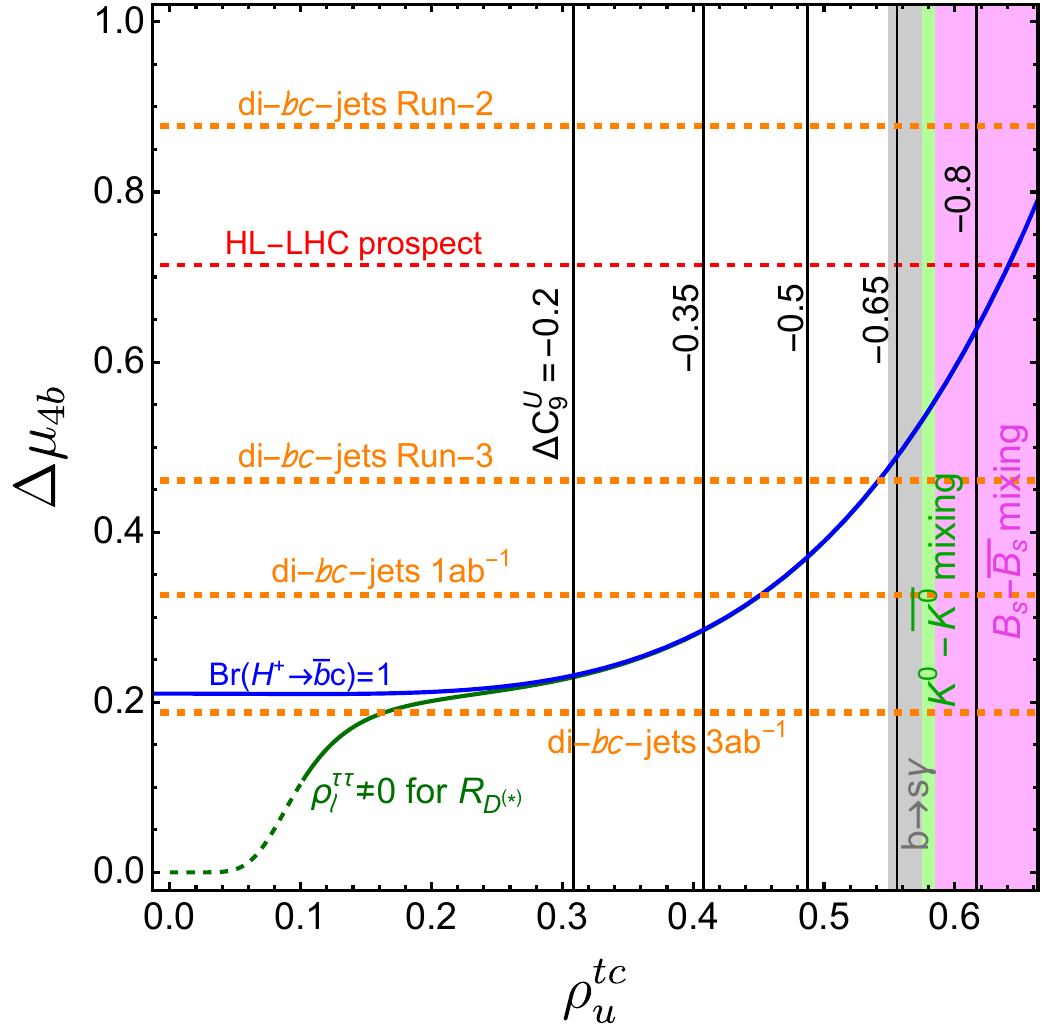}
~~~~~~%
\includegraphics[width=0.44\textwidth]{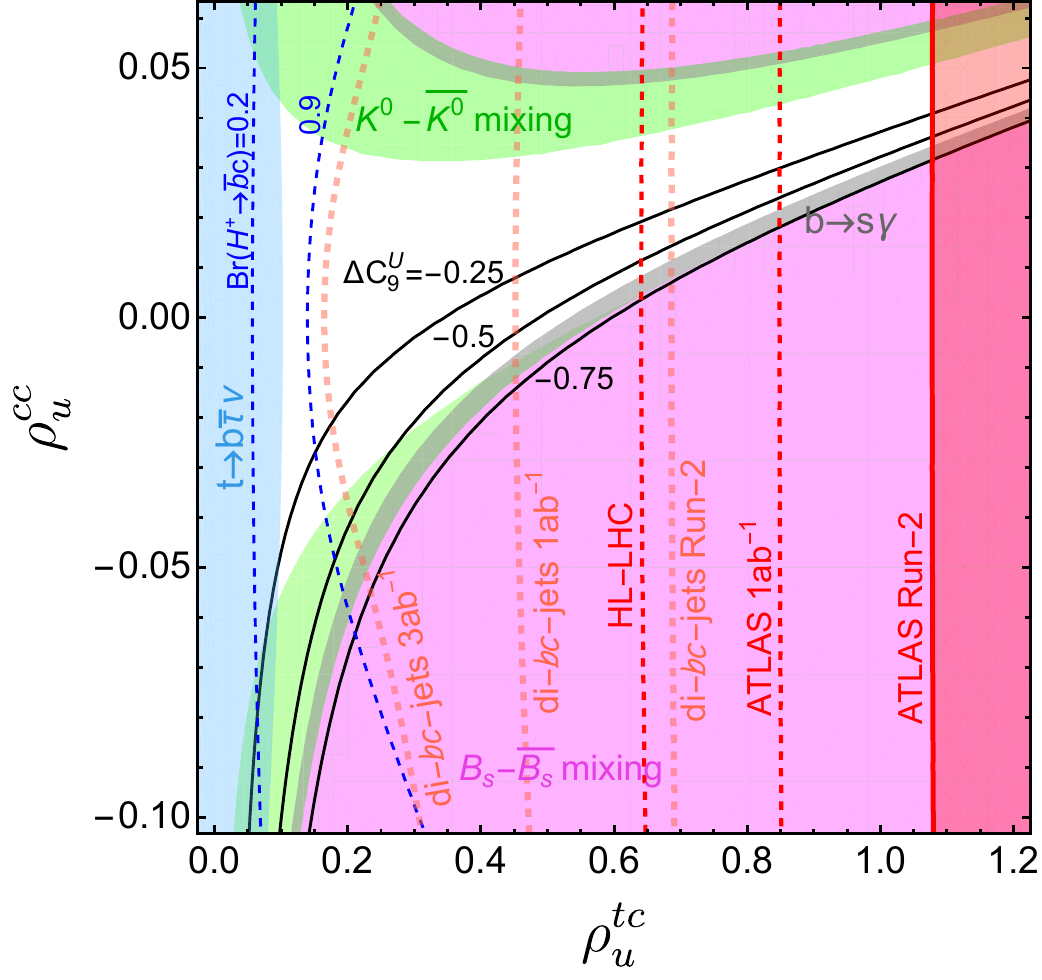}
\caption{ \justifying 
 {\bf{\it{Left:\,}}} Contours for $\Delta C^U_9$ (black vertical lines) in the $\Delta\mu_{4b}$--$\rho_u^{tc}$ plane for $\rho_u^{cc}=0$. Both the prediction for Br$(H^+\to \ov b c)=1$ (blue) and $\rho_\ell^{\tau\tau}$ chosen such that $R_{D^{(*)}}$ is explained within $1\sigma$ (green) are shown. The regions on the right are excluded by flavour constraints. The dashed part of the green line is excluded by the $t\to b\tau\ov\nu$ search at the LHC. The dashed red line shows the expected sensitivity of SM di-Higgs searches at the HL-LHC and the orange lines the expected sensitivities of dedicated searches (di-$bc$-jets).
{\bf{\it{Right:\,}}} Same observables in the $\rho_u^{cc}$--$\rho_u^{tc}$ plane with $\rho^{\tau\tau}_\ell$ fixed such that $R_{D^{(*)}}$ is explained.
 }
    \label{fig:result}
\end{figure*}

\subsection{Numerical analysis}

The benchmark (BM) points BM1 and BM3 defined in Ref.\,\cite{Crivellin:2023sig}, with $\rho_u^{tc}=0.55$ and $0.47$, predict $\mu_{4b}=3.72$ and $\mu_{4b}=2.92$, respectively, if $\epsilon_{\rm SR}=\epsilon_{\rm Loop}=\epsilon_{\rm Tree}=1$. However, due to the details of the analysis, the effect in the SM di-Higgs search turns out to be smaller, and thus the resulting bounds are weaker. This is illustrated in Fig.\,\ref{fig:result} where $\Delta\mu_{4b}= \mu_{4b} -1$ is introduced. In the left plot, $\rho_\ell^{\tau\tau}$ is set to zero for the blue line, but for the green contour $\rho_\ell^{\tau\tau}$ is fixed to explain $R_{D^{(*)}}$\footnote{More precisely $R_D=0.350$ and $R_{D^{*}}=0.271$ are obtained to explained the measurements within the $1\sigma$ level~\cite{HFLAV2024} by fixing $\rho_\ell^{\tau\tau}$ with $\rho_u^{cc}\neq 0$ scenario). More precisely, we have Br$(H^+\to \ov{b}c)>0.92$ for $\rho_u^{tc}> 0.15$ while this decreases to $50\%$ for $\rho_u^{tc}=0.08$.} and the Br$(H^+\to \ov{b}c)$ is calculated accordingly. For sufficiently small, $\rho_u^{tc}\le 0.25$, where we have $C_9^U=-0.1$ $\rho_\ell^{\tau\tau}$ needs to be larger to explain $R_{D^{(*)}}$ and Br($H^+\to \ov b c$) will thus be smaller. As a result, the corresponding $\Delta\mu_{4b}$ prediction (green) starts to deviate from the blue line and further decreases for small $\rho_u^{tc}$. Since $\rho_u^{tc}\lesssim 0.1$ is excluded by (recasted) stau searches~\cite{CMS:2022syk,Iguro:2022tmr} and $t\to b H^+\to b\ov\tau\nu$~\cite{ATLAS:2024hya}, the part of the line is depicted dashed. The $b\to s\gamma$ constraint excludes the grey region. Note that $B_s-\bar B_s$ mixing (purple) and $K^0-\bar K^0$ mixing (light green) provide slightly weaker and comparable bounds~\cite{Crivellin:2023sig}. Black vertical lines show the size of the NP contribution to $C_9^U$. The expected HL-LHC sensitivity of $\mu_{4b}$ at $2\sigma$ is shown with the red dashed line. Thus, current flavour constraints are more stringent and the HL-LHC will start to probe the interesting region $\Delta C_9^U\approx-0.5$. Similarly in the right plot of Fig.\,\ref{fig:result} we show the excluded parameter region in the $\rho_u^{cc}$--$\rho_u^{tc}$ plane. We see that the constraints from recasting SM di-Higgs searches already probe new regions in parameter space where flavor constraints are milder due to a non-zero $\rho_u^{cc}$.\footnote{We note that low mass di-jet searches at LHC give $|\rho_u^{tc}|\lesssim 1.5 $ \cite{Desai:2022zig} and thus the constraint provides the best limit.}

To enhance the sensitivity, charm tagging and QCD jet rejection are key elements. ATLAS and CMS are working on the improvements based on \texttt{Transformers} \cite{Vaswani:2017lxt,Flavor_Tagger}. However, enhanced flavor discrimination will lead not only to a larger $\epsilon_b$ but also to a smaller $\epsilon_{c\to b}$. Thus, a dedicated analysis would be very welcome and promises to have discovery potential at the LHC. First of all, searching for $c$-tagged jets has a three times higher efficiency than the $c\to b$ mis-tag probability~\cite{Sarkar:2024vjz}. Furthermore, the smallness of $\epsilon_{\rm Tree,Loop}$ and $\epsilon_{\rm SR}$ suggests that the signal is relevantly different from the SM one, such that a combined improvement by a factor $\approx 20$ in statistics could be expected for both experiments. This estimate for the $2\sigma$ sensitivity of a dedicated search is shown in Fig.~\ref{fig:result} (orange dashed line). As one can see, a dedicated search with HL-LHC data covers nearly the whole interesting region of parameter space, highlighting the discovery potential.

\section{Conclusions}

A light charged Higgs boson ($m_{H^\pm}\approx 130$\,GeV) is suggested by the ATLAS excess in $t\to b\bar bc$ ($3\sigma$) and has the potential to explain the $B$ anomalies ($R_{D^{(*)}}$ and $b\to s\ell^+\ell^-$ data) within the two-Higgs-doublet model with generic Yukawa couplings. In fact, the coupling pattern favoured by the $B$ anomalies suggests a dominant decay of $H^+$ to $\bar b c$, in agreement with the $t\to b\bar bc$ excess and the upper limit from the charged Higgs search in $t\to b \ov \tau \nu$.

In this letter, we pointed out that this scenario can be tested by recasting analyses for the pair production of SM Higgses in the $4b$ channel. This is possible because we 1) have a relevant $H^\pm$ pair-production cross section induced by the $SU(2)_L$ charge of the second Higgs doublet and its Yukawa coupling, necessary to explain the anomalies 2) the subsequent $H^+\to \ov b c$ decay has overlap with the SM signal due to the non-negligible $c$-jet to $b$-jet mistagging rate and the limited LHC-detector resolution in hadronic channels. 

However, the rather small efficiencies of the NP signal with respect to the SM one, as seen by recasting the ATLAS analysis~\cite{ATLAS:2023qzf}, reduce the effect. Nonetheless, this can be circumvented by dedicated analysis, which clearly offers the possibility of discovering charged Higgs pair production. In fact, already Run-3 data can cover a large part of the interesting parameter space, motivated by an explanation of the ATLAS excess and the $B$ anomalies, and the HL-LHC will further improve the sensitivity.

\begin{acknowledgments} 
We are very grateful to Yasuyuki Horii, Shigeki Hirose, and Hiroyasu Yonaha for enlightening discussions and encouraging this work. We also appreciate Hantian Zhang for his participation in the early stage of this work. The work of A.C.~is supported by a professorship grant from the Swiss National Science Foundation (No.\ PP00P21\_76884). S.I. is supported by the JSPS Grant-in-Aid for Scientific Research Grant No.\,22K21347, 24K22879, 24K23939, 25K17385, Core-to-Core Program Grant No.\,JPJSCCA20200002 and the Toyoaki Scholarship Foundation.
\end{acknowledgments}

\bibliographystyle{utphys}
\bibliography{references}
\end{document}